\documentclass[runningheads,a4paper]{llncs}
\usepackage[margin=2cm]{geometry}

\usepackage{epstopdf}
\usepackage{url}

\urldef{\mailsa}\path|{xiang.zhang3, manqing.dong, chang.ge}@student.unsw.edu.au,|
\urldef{\mailsb}\path|huanliucs@gmail.com|
\urldef{\mailsc}\path|{xiaocong.chen, lina.yao}@unsw.edu.au|

\usepackage{cite}
\usepackage{amsmath,amssymb,amsfonts}
\usepackage{algorithmic}
\usepackage{graphicx}
\usepackage{textcomp}
\usepackage{fancyhdr}\def\BibTeX{{\rm B\kern-.05em{\sc i\kern-.025em b}\kern-.08em
    T\kern-.1667em\lower.7ex\hbox{E}\kern-.125emX}}
\usepackage{booktabs} % For formal tables
\usepackage{colortbl}
\usepackage[ruled, vlined, linesnumbered]{algorithm2e}

\usepackage{comment}
\usepackage{mathtools}
\usepackage{multirow}
\usepackage[dvipsnames]{xcolor}
\usepackage{balance}
\usepackage{bm}
\usepackage{caption}
\usepackage{subcaption}
\usepackage{xcolor}

\usepackage{eucal}

\makeatletter
\def\endthebibliography{%
  \def\@noitemerr{\@latex@warning{Empty `thebibliography' environment}}%
  \endlist
}
\makeatother

\begin{document}

\mainmatter  % start of an individual contribution
\title{Multi-task Generative Adversarial Learning on Geometrical Shape Reconstruction from EEG Brain Signals}
\titlerunning{Geometrical Shape Reconstruction from EEG Brain Signals}

%author: Xiang, xiaocong, Lina, Chang, Huan Liu, Manqing
\author{Xiang Zhang$^\ddag$, Xiaocong Chen$^\ddag$,  Manqing Dong$^\ddag$,\\ Huan Liu$^\S$,Chang Ge$^\ddag$,  Lina Yao$^\ddag$ }
\authorrunning{Xiang Zhang et.al}
\institute{$^\ddag$ University of New South Wales, Sydney, Australia\\
$^\S$ Xi'an Jiaotong University, Xi'an, China\\
\mailsa\\ 
\mailsb\\
\mailsc\\
} 

\maketitle
\begin{abstract}
Synthesizing geometrical shapes from human brain activities is an interesting and meaningful but very challenging topic. Recently, the advancements of deep generative models like Generative Adversarial Networks (GANs) have supported the object generation from neurological signals. However, the Electroencephalograph (EEG)-based shape generation still suffer from the low realism problem. In particular, the generated geometrical shapes lack clear edges and fail to contain necessary details. In light of this, we propose a novel multi-task generative adversarial network to convert the individual's EEG signals evoked by geometrical shapes to the original geometry.
% we propose an effective framework to convert the individual's EEG signals evoked by geometrical shapes to the original geometry. 
First, we adopt a Convolutional Neural Network (CNN) to learn highly informative latent representation for the raw EEG signals, which is vital for the subsequent shape reconstruction. Next, we build the discriminator based on multi-task learning to distinguish and classify fake samples simultaneously, where the mutual promotion between different tasks improves the quality of the recovered shapes. Then, we propose a semantic alignment constraint in order to force the synthesized samples to approach the real ones in pixel-level, thus producing more compelling shapes. The proposed approach is evaluated over a local dataset and the results show that our model outperforms the competitive state-of-the-art baselines.
 
\end{abstract}

\section{Introduction}
Since the advent of neuroscience and brain-computer interface (BCI), numerous studies tried to recover the visual stimuli based on the informative human brain activities \cite{seeliger2018generative,zhang2019survey}. The development of the decoding technologies of chaotic brain signals is supposed to reveal the mechanism of brain neurons and may implement some fantastic ambitions such as mind reading \cite{zhang2018converting}. 
Most of the existing work focused on functional magnetic resonance imaging (fMRI) monitoring brain activities by detecting changes associated with blood flow in brain areas. However, fMRI-based image reconstruction faces several major challenges \cite{nishimoto2011reconstructing,seeliger2018generative}. The temporal resolution of fMRI is low constrained by the blood flow speed; the acquisition of fMRI requires a scanner which is expensive and hard to afford; the scanner is heavy and has poor portability \cite{zhang2019survey}. 

\begin{figure}[t!]
\centering
    \includegraphics[width=0.8\linewidth]{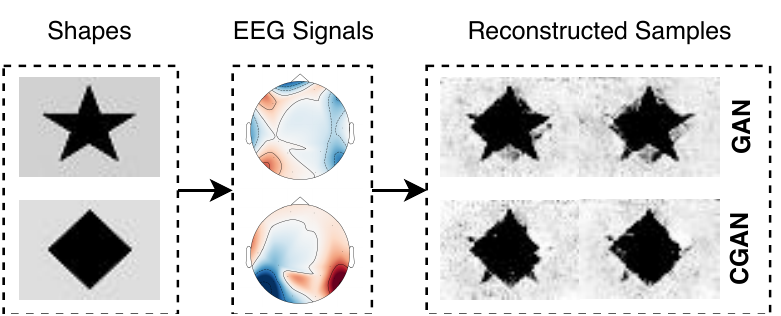}
    \caption{Generated samples based on EEG signals evoked by geometric shapes. It is observed that the samples synthesized by traditional methods (e.g., GAN and CGAN) are blur and lack of realistic details.}
    \label{fig:introduction}
    % \vspace{-3mm}
\end{figure}

Thus, Electroencephalogram (EEG) recently has drawn much attention as its high temporal resolution, low price, and high portability. EEG is a non-invasive signal measuring the voltage fluctuations generated by an electrical current within human neurons. Researchers have tried to exploit EEG signals to reconstruct visual stimuli \cite{kavasidis2017brain2image,palazzo2017generative} through Generative Adversarial Networks (GANs). Nevertheless, the previous studies suffer from the low realism problem of the generated samples, which means that the model can not generate images with high realism based on the input brain signals. In other words, the current EEG-based synthesis methods can roughly present the visual stimuli but fail to contain necessary details. For example, as shown in Figure~\ref{fig:introduction}, the clear geometric shapes are present to the individual and reconstruct the shapes based on the collected EEG data. It is demonstrated that the geometric shapes generated by traditional GAN and CGAN are blurry and lack of realistic details. 

\begin{comment}
Additionally, the previous works are mainly focused on the reconstruction of the image of objects (e.g., car and plane) which contain too many attributes (e.g., color, shape, size, background, and semantic information), as a result, it is difficult to figure out the human brain is more sensitive to which attribute and which one contributes more to the object reconstruction. \textcolor{red}{To this end, the investigation on the sensitivity of the brain to a specific attribute (e.g., geometric shape) is meaningful and necessary.} 
\end{comment}

Aiming at the aforementioned issues, in this paper, we conduct experiments to measure the individual's EEG oscillation evoked by various geometrical shapes and propose a novel framework in order to precisely decode the EEG signals and synthesize the geometric shapes.
Moreover, we employ a Convolutional Neural Network (CNNs) to explore the latent representation form the raw EEG signals since CNN is much efficient than the Recurrent Neural Networks (RNNs) with a similar EEG representation learning ability based on our empirical experiments. In addition, we adopted a multi-task discriminator with a task-specific classifier which assigns the geometric shape into the correct class for the aim of improving the quality of the recovered shapes. Furthermore, we propose a semantic alignment method involving the semantic information of the real shape to enhance the realism level of the reconstructed shape. The previous works are mainly paid attention to brain signal based images (e.g., bird and plane) reconstruction which contain too many attributes (e.g., color, shape, size, background, and semantic information), as a result, it is difficult to figure out which attribute the human brain is more sensitive to and which one contributes more to the object reconstruction. Thus, in this work, we focus on the EEG-based geometric shape reconstruction and attempt to illustrate that EEG signals are sensitive to geometries. 
% I think illustrate should followed by some sentence like "EEG signals is sensitive to geometries" instead of "is...." if we are going to use "is EEG...." it could be better to write it as "attempt to find out is EEG signals sensitive......"?

In detail, the contributions of this work are listed here:
\begin{itemize}
    \item We present a novel deep generative model to recover the geometrical shape seen by human beings from the EEG signals. To our best knowledge, we are the first work investigating the brain signal based geometric shape reconstruction. The reproducible codes are publicly available here\footnote{https://github.com/xiangzhang1015/EEG\_\_Shape\_Reconstruction}.

    \item We propose an effective semantic alignment method to harness the semantic information of the original geometric shape in order to force the approach to produce more realistic shapes.

    \item We conducted a local EEG dataset stimulated by various geometric shapes and evaluate the proposed approach over the collected dataset. The experimental results demonstrated that our model outperforms all the competitive state-of-the-art baselines.
\end{itemize}

\section{Related Work} % (fold)
\label{sec:related_work}
Recent years' research in neuroscience and neuroimaging \cite{haynes2006neuroimaging} indicated that human perception of visual stimuli can be decoded through some techniques in neuroimaging. To be specific, a few works gave evidence about decoding the brain signals to human activity by using the Functional Magnetic Resonance Imaging (fMRI) and EEG. There are some works use the fMRI signals to reconstruct the image which is seen by the individual and get an acceptable performance \cite{nishimoto2011reconstructing,naselaris2009bayesian}. The studies show the potential of fMRI-based image reconstruction in the brain signals decoding area, however, fMRI faces a number of crucial issues such as expensive acquisition equipment and low portability. 
% but the limitation of fMRI is very obvious which is the high experimental cost. 
Apart from the fMRI based method, there are a few EEG based methods in image reconstruction as EEG signals are less expensive \cite{kavasidis2017brain2image,palazzo2017generative}. 
As a typical investigation, Brain2image \cite{kavasidis2017brain2image} encoded the raw EEG signals into a latent space which contains the distinctive information, and then sent them to a Conditional Generative Adversarial Networks (CGAN) for image reconstruction. Palazzo et al. \cite{palazzo2017generative} applied a very similar algorithm framework. 

Most of the visual object reconstruction methods are based on Generative Adversarial Networks (GANs) and the variations. 
GANs \cite{goodfellow2014generative}, as the typical deep learning frameworks, was used widely in image generation. The standard GANs are composed of a generator network which generates images from the random sampled noise and a discriminator network which tried to distinguish the generated image correctly. Normally, original GANs had to suffer from the uncontrollable issue of the generation process. In order to retard it, the conditional GAN (CGAN) was proposed \cite{mirza2014conditional} which involves the conditional information (e.g., labels) in order to control the generating process. Auxiliary Classifier GAN (ACGAN)\cite{odena2017conditional} improve the performance of GAN for image synthesis. ACGAN demonstrated that adding more structure to the GAN latent space along with a specialized cost function results in higher quality samples. A task-specific branch in the discriminator is empowered to enhance the discriminability.

\noindent\textbf{Summary.} Most brain signal based image reconstruction work is based on fMRI. Due to the drawbacks of fMRI (e.g., low time resolution, expensive, and low portability), we focus on EEG based geometric shape reconstruction. Compare to the typical EEG-based work like brain2image \cite{kavasidis2017brain2image}, we have several technical advantages: 1) we concentrate on the influence to the EEG signals brought by geometric attribute while \cite{kavasidis2017brain2image} focus on images with a large number of attributes; 2) we adopt CNN instead of RNN to learn the latent EEG features which cost less training time with a similar accuracy; 3) we add an auxiliary task-specific classifier to improve the discriminability of the discriminator; 4) we propose a semantic alignment method to generate more realistic images.

\begin{figure}[t!]
    \includegraphics[width=\linewidth]{CNN.png}
    \caption{Demonstration of discriminative EEG representation learning. The last second layer $\bm{\bar{E}}$ with discriminative information is selected as learned representation. Each Conv stage contains a convolutional layer followed by a pooling layer. The basic hyper-parameters are presented.} 
    \label{fig:cnn}
    % \vspace{-3mm}
\end{figure}

\section{Method} 
\label{sec:method}
In this study, we aim to propose a method to convert the individual's mental geometry into physical shape. In particular, we first decode the non-invasive EEG signals into an implicit representation (Section~\ref{sub:discriminative_representation_learning}) and then propose a modified GAN framework to generate the real shape which evoked the EEG signals (Section~\ref{sub:image_generation_model}. 
In this section, we will introduce the workflow of the whole system in detail.

\subsection{EEG Feature Learning} % (fold)
\label{sub:discriminative_representation_learning} 

In the EEG feature learning, we adopt a CNN structure to capture the latent distinguishable features from the collected EEG signals. Some research had demonstrated that CNN is empowered to learn informative features from noisy EEG data\cite{zhang2019vulnerability,acharya2018deep}. Suppose the EEG sample pairs can be denoted by $\bm{E}=\{ (\bm{E}_h, \bm{y}_h), h =0, 1, \cdots H\}$ where $\bm{E}_h \in \mathbb{R}^{M \times N}$ and $\bm{y}_h \in \mathbb{R}^5$ represent the EEG observations and the corresponding one-hot label. In this paper, we focused on the decoding of five different visual-stimuli evoked imagination, thus the number of labels is five. 
The $H$ denotes the number of EEG segments and $M, N$ denotes the time- and spatial- resolution of each segment. 

Figure \ref{fig:cnn} shows the workflow of the learning procedure of the discriminative representation. The visual-stimuli evoked EEG signals, reflecting the imagination in the user's mind, are feed into a CNN model with seven layers. The first convolutional layer contains 32 filters with the kernel size of $[3, 3]$ and stride of $[1, 1]$. The padding method is `SAME' while the activation function is ReLU. The first pooling layer adopts max pooling and both the pooling size and strides are $[2, 2]$. The second convolutional and pooling layers are identical to the first layers, respectively, except the Conv 2 has 64 filters. The followed fully-connected layer has $d$ nodes, which is regarded as the learned representation, denoted by $\bm{\bar{E}}$,  and contains enough information to reconstruct the visual shape. The learning algorithm iterates for 1,000 epochs with Adam optimizer has a learning rate of $5 e^{-4}$. 
% The original EEG signals were fed into several convolutions and pooling layers which the output will go to a fully-connected layer. And the representation of this fully connected layer is the representation what we want to acquire. 

Compared to Brain2Image \cite{kavasidis2017brain2image} which employed LSTM for feature learning, CNN is able to achieve a similar performance but spend much less training time. In particular, LSTM obtained the classification accuracy of 74\% with 5,935s while CNN achieved 72\% but with only $1,222$s.

\begin{figure}[t!]
    \includegraphics[width=\linewidth]{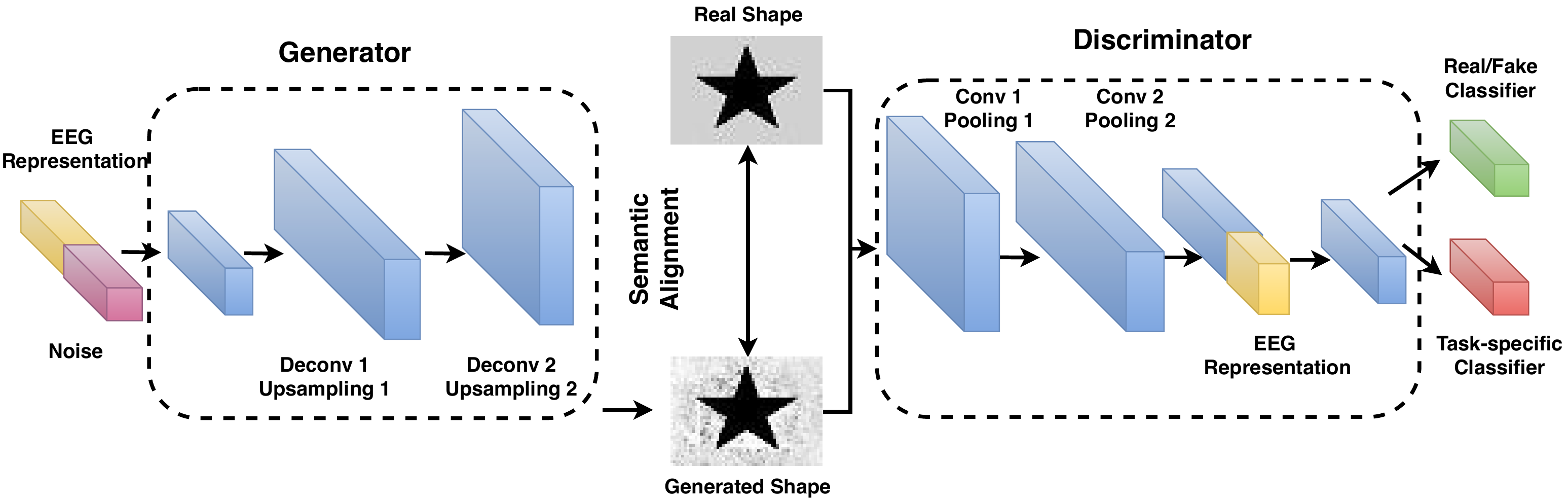}
    \caption{Workflow of the proposed visual stimuli reconstruction framework. We adopted a semantic classifier apart from the real/fake classifier in order to exploit the semantic information of the EEG samples. Moreover, a semantic regularization constraint is proposed to force the generated visual stimuli has similar semantic information with the real visual stimuli.
    } 
    \label{fig:workflow}
    % \vspace{-3mm}
\end{figure}

\subsection{Multi-task Generation Model} % (fold)
\label{sub:image_generation_model}

\subsubsection{Overview} % (fold)
\label{sub:overview}

% subsection overview (end)
In this part, we will describe the framework which is used to reconstruct the shapes that human seeing. As shown in Figure~\ref{fig:workflow}, the proposed geometrical shape generation framework contains two components: a generator and a discriminator. 

The generator receives the learned discriminative EEG representation $\bm{\bar{E}} \in \mathbb{R}^d$ along with a random sampled Gaussian noise $\bm{z} \in \mathbb{R}^{d'}$ and produces generated shape. The EEG representation is evolved to guarantee the compelling of the generated shapes while the Gaussian noise is adopted to keep the diversity. 
% In short, this method can force the generator to produce a certain shape corresponding to 
On the other hand, the discriminator receives the real shape which evoked the brain signals (the imagination which presented in the human brain) and the generated fake shape. Inspired by ACGAN \cite{odena2017conditional}, we design a multi-task discriminator containing two branches while the first branch, like the standard GAN, aims at the recognition of the fake shapes and the second branch, an auxiliary task-specific classifier, attempts to classify what class the shape belongs to. The first branch is called real/fake classifier whilst the second one is called task-specific classifier. By adding the task-specific classifier, the designed discriminator not only is able to distinguish whether the shape is real or not but also can recognize the category of the shape. As a consequence, the discriminator drives the distribution of the synthesized shapes not only approximate to the general distribution of the overall real shapes but also approximate to the distribution of a specific category. In addition, the learned EEG representation is also input to the discriminator, as proposed in \cite{mirza2014conditional}, in order to make the discriminator under the same conditional situation with the generator.

\subsubsection{Architecture} % (fold)
\label{sub:architecture}

% subsection architecture (end)
Next we report the details of the architecture. The generator receives the input vector which concatenates $\bm{\bar{E}}$ and $\bm{z}$, represented by $\bm{h_0} = \{\bm{\bar{E}}:\bm{z}\} \in \mathbb{R}^{d+d'}$, and attempts to map it to a meaningful shape. The generator is composed of a fully-connected and two deconvolutional layers each followed by a unsampling layer. The $\bm{h_0}$ is first fed into the fully-connected layer with $64(M+N)$ nodes:
\begin{equation}
    \bm{h_1} = \sigma (\bm{w}\bm{h_0} + \bm{b})
\end{equation}
where $\bm{w}, \bm{b}$ and $\sigma$ denote the weight, bias vector, and the sigmoid function, respectively.  
Then $\bm{h_1}$ is reshaped into $[M, N, 64]$ where 64 denotes the depth. To this end, $\bm{h_1}$ has a similar form, but deeper depth, with the raw EEG segment $\bm{E_h}$ which is supposed to contain enough information to reconstruct the user's imagination. Afterward, $\bm{h_1}$ is sent to the the first deconvolutional layer with 32 filters, kernel size $[5, 5]$, stride $[2,2]$, and 'SAME' padding method. The upsampling operation is the invert operation of pooling and shares the same parameters with pooling layer. The second deconvolutional with one filter and upsampling layers. We choose the tanh as activation function since it transforms the signals into the range $[-1, 1]$ which is the same range the real shape falls into. 
The synthesized shape $\bm{F}$ has shape $[4M, 4N]$. According to empirical experiments, we set the shape size 4 times of the EEG raw segment in both width and height in order to have a better generation quality. The real geometric shape $\bm{R}$ is in greyscale with format $[4M, 4N]$. All the pixels are normalized into the range $[0,1]$ by max-min normalization and then transformed to $[-1, 1]$ by:
\begin{equation}
    \bm{\bar{R}} = 2\bm{R} - 1
\end{equation}

In the discriminator, as shown in Figure~\ref{fig:workflow}, both $\bm{\bar{R}}$ and $\bm{F}$ are fed into the discriminator which has almost the same structure and hyper-parameters with the discriminative representation learning model (Section~\ref{sub:discriminative_representation_learning}). The input shape is flattened to a vector and then concatenates with the learned representation $\bm{\bar{E}}$. The fully-connected layer has 100 nodes. 
This designed discriminator has two branches corresponding two output layers. The output layer of the real/fake classifier only has one node which represents the fake probability. As for the task-specific classifier, the output layer has five nodes corresponding to five different geometrical shape categories. 

\subsubsection{Loss Function} % (fold)
\label{sub:loss_function}
We present the loss functions in the proposed framework. For the generator, since we add a task-specific classifier, the loss function contains two components where one component forces the discriminator cannot recognize the shape is generated while another component forces the discriminator to recognize which shape category the shape belongs to. Thus, the log-likelihood loss function for the generator can be defined as \cite{odena2017conditional}: 
\begin{equation}
\label{eq:g_loss}
      \mathcal{L}_g = E[\log P(C=\bm{y}|X=\bm{F})] + E[\log(1-D(G(\bm{y},\bm{\bar{E}}, \bm{z})))] %+ \lambda  S_r 
\end{equation}
in which, 
\begin{equation}
    \bm{F} = G(\bm{y},\bm{\bar{E}}, \bm{z})
\end{equation}
describes the generator $G$, and 
\begin{equation}
     P(S|X), P(C|X) = D(X)
\end{equation}
describes the real/fake classifier and task-specific classifier of the discriminator $D$, respectively.
As for the discriminator, the loss function also contains two components separately coming from the two classifiers. The discriminator is supposed to filter out which shape is generated, meanwhile, to assign the shape into the correct class. 
The log-likelihood loss function $\mathcal{L}_d$ for the discriminator is: 
\begin{equation}
        \mathcal{L}_d = E[\log P(S=\bm{\bar{R}}|X=\bm{\bar{R}})] + E[\log P(S=\bm{F}|X=\bm{F})] + E[\log P(C=\bm{y}|X=\bm{\bar{R}})]
\end{equation}
In the above formula, the $\bm{y}$ represents the class label. The $C, S$ denote the predicted class and and source, which are the classification results of the multi-task generator. $X$ denotes the shape fed into the discriminator. The $P(S|X)$ denotes the probability distribution over the source $S$ while the $P(C|X)$ denotes the probability distribution over the class label $\bm{y}$.

\subsection{Semantic Alignment} % (fold)
\label{sec:class}
To this end, the geometrical shape reconstruction model is able to generate a batch of samples which have enough diversity but still less discriminability. Furthermore, in order to increase the discriminability of the generated samples and make the samples more realistic, we propose a semantic alignment method to adopt the semantic information to make the synthesized shape more realistic and sharper. In particular, we add an additional constraint on the generator loss function aiming at reducing the distance between the real and the generated geometric shapes. 

The semantic distance can be measured by $S_r$:
\begin{equation}
        S_r = \frac{1}{\sqrt{\bar{N}}}\sqrt{\sum_{i=0}^{\bar{N}} \sum_{j=0}^{\bar{N}}(\bar{R}_{i,j} - F_{i,j})^2}
\end{equation}
where $\bar{N}$ denotes the number of pixels in the geometric sample and $\bar{N} = 4M \times 4N$. 
The $\bar{R}_{i,j}$ and $F_{i,j}$ denote the pixels in the real and generated samples.
 % is one of the pixels in the real shape, $F_{i,j}$ is one of the pixels in the fake shape. 
 In order to improve the performance of the generator, the $S_r$ is considered as a regularization of the generator loss. Thus, we update the Equation~\ref{eq:g_loss} as:
\begin{equation}
      \mathcal{L}_g = E[\log P(C=\bm{y}|X=\bm{F})] + E[\log(1-D(G(\bm{y},\bm{\bar{E}}, \bm{z})))] + \lambda  S_r 
\end{equation}
where $\lambda$ is a constant coefficient to adjust the weight of semantic regularization. If the alignment constraint too strong, the generated shapes may have less diversity. In this work, we set $\lambda=0.01$ to make a trade-off between the diversity and discriminability of the generated samples.
% which is set as 0.01. 

During the training, both $\mathcal{L}_g$ and $\mathcal{L}_d$ are optimized by the Adam optimizer. The learning rate is set as 0.0002 with the exponential decay rate of 0.5. In each epoch, the $\mathcal{L}_g$ and $\mathcal{L}_d$ are separately trained in turn. The proposed framework converges after 120 epochs and trend to overfitting after 160 epochs, thus, we adopt the early stopping strategy by breaking the iteration at the 150-th epoch.

\section{Experiments} % (fold)
\label{sec:experiments}
In this section, we will describe the experiments and the performance analysis containing qualitative and quantitative aspects in detail. The qualitative comparison will conduct the analysis in the quality of the generated shapes, and the quantitative comparison will be based on the inception score \cite{salimans2016improved} and inception accuracy.

\subsection{EEG Signal Acquisition} % (fold)
\label{sub:brain_signal_acquisition} 
We conducted a local experiment with 8 healthy participants (6 males and 2 females) aged 25 $\pm$ 3, which is approved by UNSW ethic abroad (HC190315). During the experiments, the participant is required to sit in an armed comfortable chair in front of a computer monitor. We select five representative and widely-seen geometrical shapes (circle, star, triangle, rhombus, and rectangle) to present to the subject. The whole experiments contain two sessions and each session has five trials. In each trial, the five geometrical shapes are presented in random order and each shape lasts for five seconds. There are five seconds relax period among two adjacent shapes. The relaxing time among trials and sessions are 10s and 30s, respectively. The EEG signals are collected through a portable Emotiv EPOC+ headset with 14 electrodes and the sampling frequency is set as 128 Hz. Each EEG segment contains ten continuous instances with 50\% overlapping. The dataset is randomly divided into a training set (80\% proportion) and testing set (20\% proportion).  

Based on the collected EEG data, we report the hyper-parameters settings. The single EEG segment $\bm{E}$ ( $M = 10$ and $N=14$) is compressed into a latent discriminative representation $\bm{\bar{E}}$ with dimension $d=40$. In the generator, the stochastically sampled noise $\bm{z}$ has dimension $d'=20$. The coefficient of semantic regularization $\lambda$ is set as 0.001.

\begin{figure}[t]
    \includegraphics[width=\linewidth]{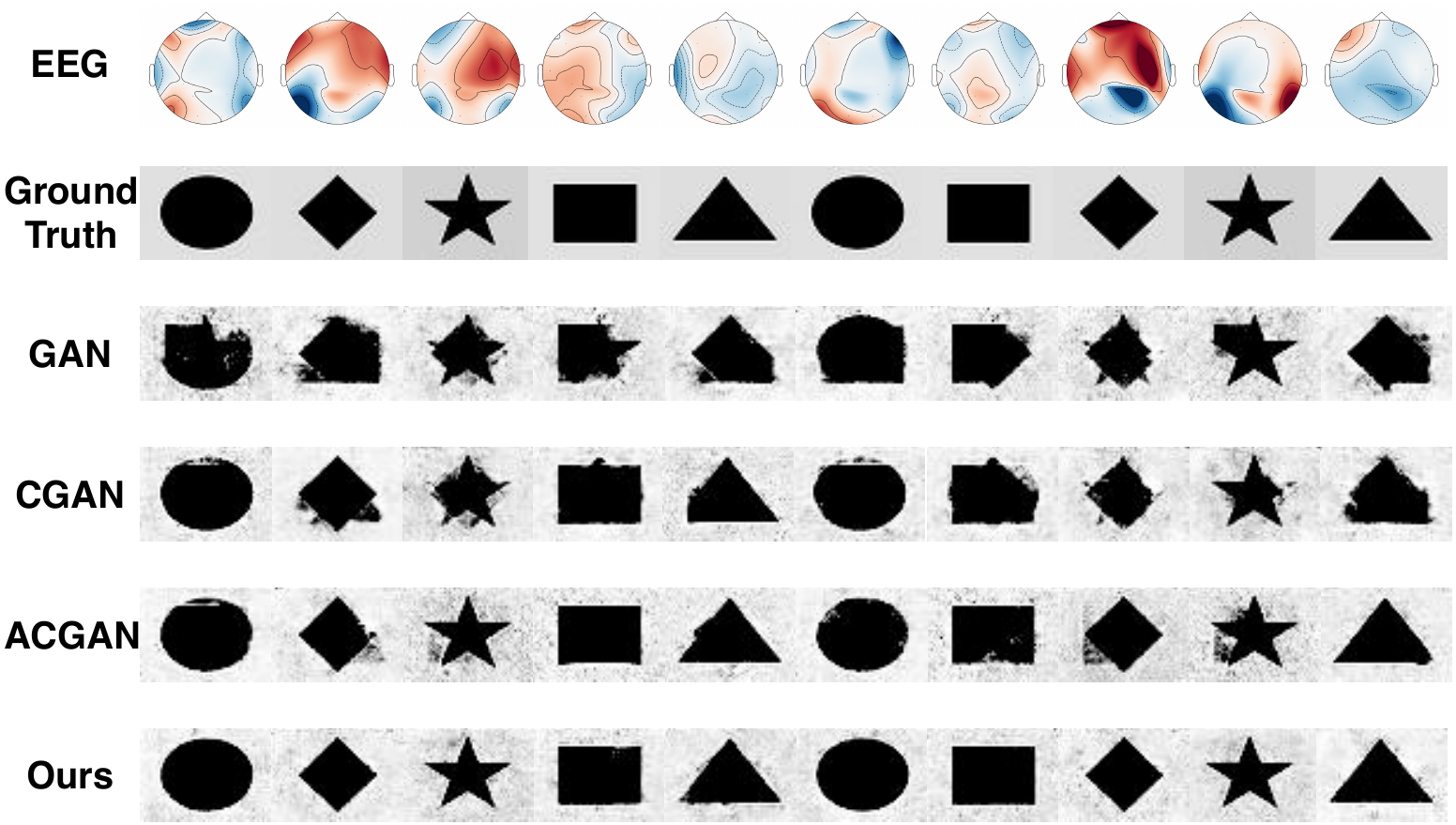}
    \caption{Demonstration of the qualitative comparison. Our model can reconstruct all the shapes correctly which have the highest similarity with the ground truth.} 
    \label{fig:comparison}
    % \vspace{-3mm}
\end{figure}
\subsection{Qualitative Comparison} % (fold)
\label{sub:comparison}
In this section, we compare the quality of the generated shapes among the proposed method and the state-of-the-art models. As shown in Figure \ref{fig:comparison}, we choose the most widely used generative models including GAN, CGAN and ACGAN as the baseline.

GAN achieve a promising result in many areas, especially in shape field \cite{goodfellow2014generative}. On the top of basic GAN, CGAN \cite{mirza2014conditional} is proposed to add the conditional information as a constraint, which is adopted in \cite{kavasidis2017brain2image}. Furthermore, ACGAN attempt to deeply exploit the informative sample labels to enhance the discriminability of $D$ \cite{odena2017conditional}. Our work, compared to ACGAN, proposed a semantic alignment method to constrain the distance among the synthesized shapes and the visual geometrical shapes in order to further emphasize the reality.

It's easy to find that, from Figure \ref{fig:comparison}, our approach have the best shape quality. To be specific, the samples which generated by GAN are lack of clear edge, which is a typical mode collapse problem, meanwhile, it's not hard to figure out that most of the synthesized shapes have miscellaneous features. The CGAN has a better performance than normal GAN as the shapes have a higher integrity. However, we still can find that some shapes generated by CGAN have combined features such as a star have the feature from rhombus. The ACGAN have the best result among the baseline models, which it can learn most of the shapes' feature and correctly reconstruct the shapes with a trivial acceptable flaw. Our model can reconstruct all the shapes correctly which have the highest similarity with the ground truth.

\subsection{Quantitative Comparison} % (fold)
\label{sub:quantitative_comparison}
The qualitative comparison is relatively easy as the shape quality is the assessment criteria. The quantitative analyses are hard to conduct as the comparison between reconstructed and real shape is not obvious and clearly defined. The common way we used to do that is using the inception score and the inception accuracy \cite{kavasidis2017brain2image}. We build an inception network used the generated shapes as input in order to calculate the inception score which measures how realistic the generated shapes are. In detail, we generate 1,000 images for each geometric shape and calculate the overall inception score.
Moreover, our work is supposed to convert the specific EEG signals into the corresponding geometrical shape belonging to the specific label. Thus, we adopt the performance of the task-specific classifier when the input data is $\bm{F}$ as inception accuracy in order to measure how precise can our model generates shapes.

We conduct the quantitative analyses for the baselines and our proposed model. The results are presented in Table \ref{tab:inception}, in which, it is easy to observe that our model achieves the highest inception score and inception accuracy of 2.178 and 0.83, respectively. 
The inception score is not good as the public datasets like CIFAR-10 and the most possible reason is that our generated shapes are conditioned by EEG signals which is chaotic and has a low signal-to-noise ratio.
Even though, the proposed approach outperforms all the competitive baselines.

\begin{table}[t]
    \centering
    \caption{The quantitative comparison of inception score and inception accuracy}
    % \vspace{3mm}
    \label{tab:inception}
    \resizebox{0.7\textwidth}{!}{
    \begin{tabular}{l|l|l|l|l}
    \hline
    \textbf{Models} & \textbf{GAN} & \textbf{C-GAN \cite{kavasidis2017brain2image}} & \textbf{ACGAN} & \textbf{Ours} \\ \hline
    Inception Score & 1.931        & 1.986          & 2.061          & \textbf{2.178}      \\
    Inception Accuracy & 0.43        &0.67          & 0.79          & \textbf{0.83}      \\ \hline  
    \end{tabular}
    }
\end{table}

\section{Discussion and Future Work} % (fold)
\label{sec:discussion_and_futurework}
In this section, we discuss the opening challenges and potential future work of our research. 

First of all, one major issue faced by brain signal based reconstruction is the recovery of unseen geometrical shapes. For instance, one future scope is to decode the EEG signals evoked by star while the star never is trained in the reconstruction model. One possible solution is train a common generative model by a large classes of basic geometrical shapes (e.g., circle, ellipse, straight line, triangle, rectangle, and rhombus) in order to learn the latent features of each different shape and then approximate the unseen shape (e.g., star) based on the learned features.

Second, we only focused on the simple geometrical shapes in this work, as a preliminary study, however, the real world application demands more complex shapes like a bow. One of our future works is to consider more complicated geometric shapes in the experiments. In addition, another potential research direction is to increase the number of geometrical categories since this work only evaluated five basic classes.

Last but not least, more participants should be involved in the experiments in order to provide a general generative model which is robust for different individuals. The influence of inter-subject divergence should be taken into account in future research.

\section{Conclusion} % (fold)
\label{sec:discussion_and_conclusion}

In this paper, we propose a novel approach to reconstruct the geometrical shape based on the brain signals. We first develop a framework learning the latent discriminative representation of the raw EEG signals, and then, based on the learned representation, we propose an adversarial reconstruction framework to recover the geometric shapes which are visualizing by the human. In particular, we propose a semantic alignment method to improve the realism of the generated samples and force the framework to generate more realistic geometric shapes. The proposed approach is evaluated over a local dataset and the experiments show that our model outperforms the competitive state-of-the-art methods both quantitatively and qualitatively.

\bibliographystyle{splncs03}
\bibliography{shape}
\end{document}